\documentstyle[draft,epsfig]{mn}
 
\def\simgt{\mathrel{\raise0.35ex\hbox{$\scriptstyle >$}\kern-0.6em
\lower0.40ex\hbox{{$\scriptstyle \sim$}}}}
\def\simlt{\mathrel{\raise0.35ex\hbox{$\scriptstyle <$}\kern-0.6em
\lower0.40ex\hbox{{$\scriptstyle \sim$}}}}

\newcommand{\si}{\sigma}
\newcommand{\Dc}{\Delta_{\rm c}}
\newcommand{\omm}{\Omega_{\rm M}}
\newcommand{\oml}{\Omega_{\Lambda}}
\newcommand{\Ga}{\Gamma}
\newcommand{\ASCA}{{\sl ASCA}}
\newcommand{\ROSAT}{{\sl ROSAT}}
\newcommand{\Einstein}{{\sl Einstein}}
\newcommand{\BeppoSAX}{{\sl BeppoSAX}}
\newcommand{\Chandra}{{\sl Chandra}}
\newcommand{\XMM}{{\sl XMM-Newton}}
\newcommand{\MAP}{{\sl MAP}}
\newcommand{\pho}{\phantom{1}}

\input epsf
\input psfig.sty
\input epsfig.sty

\begin{document}
 
\journal{Preprint UBC-COS-00-05, astro-ph/0010039}
\title[Cluster normalization]
{Power spectrum normalization from the local abundance of rich
clusters of galaxies}

\author[E.~Pierpaoli, D.~Scott \& M.~White]
{Elena Pierpaoli${}^1$, Douglas Scott${}^1$ and Martin White${}^2$\\
${}^1$Department of Physics and Astronomy, University of British Columbia,
BC V6T1Z1~~Canada\\
${}^2$Harvard-Smithsonian Center for Astrophysics, Cambridge, MA 02138}

\date{Accepted ... ;
      Received ... ;
      in original form ...}

\pagerange{000--000}

\maketitle

\begin{abstract}
The number density of rich galaxy clusters still provides the most robust
way of normalizing the power spectrum of dark matter perturbations on scales
relevant to large-scale structure.
We revisit this constraint in light of several recent developments:
(1) the availability of well-defined samples of local clusters with relatively
accurate X-ray temperatures;
(2) new theoretical mass functions for dark matter haloes which provide a good
fit to large numerical simulations;
(3) more accurate mass-temperature relations from larger catalogues of
hydrodynamical simulations;
(4) the requirement to consider closed as well as open and flat cosmologies
to obtain full multi-parameter likelihood constraints for CMB and SNe studies.
We present a new sample of clusters drawn from the literature and use this
sample to obtain improved results on $\sigma_8$, the normalization of the
matter power spectrum on scales of $8\,h^{-1}\,$Mpc, as a function of the
matter density and cosmological constant in a Universe with general curvature.
We discuss our differences with previous work, and the remaining major
sources of uncertainty.  Final results on the normalization,
approximately independent of power spectrum shape, can be expressed as 
constraints on $\si$ at an appropriate cluster normalization scale
$R_{\rm Cl}$.  We provide fitting formulas for $R_{\rm Cl}$ and
$\si(R_{\rm Cl})$ for general cosmologies, as 
well as for $\sigma_8$ as a function of cosmology and shape parameter $\Gamma$.
For flat models we find approximately
$\sigma_8\simeq (0.495^{+0.034}_{-0.037})\,\Omega_{\rm M}^{-0.60}$ 
for $\Gamma=0.23$, where the
error bar is dominated by uncertainty in the mass-temperature relation.
\end{abstract}


\section{Introduction}

In theories of hierarchical structure formation the class of objects most
recently formed holds a special significance.  Observationally this class
of objects is clusters of galaxies -- the largest virialized structures
in the present day universe.
The local abundance of rich clusters of galaxies provides a strong constraint
on the fluctuations in the matter density on scales of order 10\,Mpc
\cite{Evr89,FWED,BonMye91,HA91,Kaiser,Lil,OukBla,BahCen,Han,WEF}.
Consistency with this constraint is one of the most important tests
a model can pass, since the constraint is directly on the linear theory
power spectrum, at a scale where there is an abundance of data.
By fixing the normalization at wavelengths much smaller than those
probed by {\sl COBE\/} one obtains an accurate local normalization on
scales relevant to much of structure formation, a long lever arm for
constraining the shape of the power spectrum, and a normalization to matter
fluctuations which is independent of galaxy bias.

There have been several recent and detailed studies of the
cluster abundance, including
Bond \& Myers \shortcite{BonMye96},
Eke, Cole \& Frenk~\shortcite{ECF},
Viana \& Liddle \shortcite{VL96,VL98},
Colafrancesco, Mazzotta \& Vittorio \shortcite{ColMV},
Kitayama \& Suto \shortcite{KitSut},
Eke et al.~\shortcite{ECFH},
Pen \shortcite{Pen},
Wang \& Steinhardt \shortcite{WanSte},
Donahue \& Voit \shortcite{DoVoi99}
and Henry \shortcite{Henry}.  However, even more recently,
there have been several developments in terms of both
theory and observation which suggest it would be useful to revisit this
constraint.
Firstly the addition of \ASCA\ temperatures
\cite{ASCA} means that there is now a well
defined local temperature function for clusters, with relatively small
errors in temperature.
Secondly a number of large N-body simulations have accurately determined the
mass function of virialized haloes
(e.g.~Governato et al.~1999), finding non-negligible deviations from the
old Press-Schechter~(1974; hereafter PS) theory.
For example the extremely large
N-body simulations of the Virgo consortium have highlighted systematic
departures from the PS predicted mass functions
\cite{JFWCCEY},
which alter the constraints on the power spectrum normalization coming from
the cluster abundance.
Thirdly, more ambitious hydrodynamical simulations of cluster formation
(e.g.~Frenk et al.~1999, and references therein) have resulted in improvements
in the relationship between mass and temperature and a better estimate of its
scatter.  Finally, the increased sophistication of multi-parameter
cosmological studies (in particular driven by recent CMB anisotropy
measurements) requires that cosmological models with general curvature be
considered (see e.g.~White \& Scott 1996).

With these refinements our results are an improvement over other studies of
the past few years.  We point out explicitly where we differ from other work,
and also where we think things could be further improved in the future.
The outline of the paper is as follows:
we review some of the appropriate theory in \S\ref{sec:theory};
our local sample of clusters is presented in \S\ref{sec:data};
we describe our statistical method in \S\ref{sec:stats}; and we present
our results and conclusions in \S\ref{sec:results} and \S\ref{sec:conclusions}.

\section{Theory} \label{sec:theory}

The abundance of rich clusters is tied to the normalization of the
power spectrum (extrapolated to the present using linear theory)
through the Press-Schechter~(1974) theory and
its extensions (see Sheth, Mo \& Tormen~2000 and references therein).
The theory has always had a somewhat weak analytic
justification, its widespread adoption arising from the dual facts that
it is easy to use and provides a remarkably good fit to more computationally
expensive simulations.

Although galaxy velocity dispersion and gravitational lensing mass
estimates exist for many clusters
\cite{Caretal,Smaetal,Gir98a,Allen,Wuetal,Gir00},
the direct estimation of mass through
determination of the X-ray temperature is far more reliable
-- although the situation is certainly improving, allowing for
estimates of the mass function \cite{Gir98b,Ste99}.
In the X-ray band there is considerably more high quality data on luminosity
than on temperature, but there is enormous uncertainty in deriving mass
$M$ from luminosity $L_{\rm X}$.  Hence
our point of comparison between theory and observation will be the temperature
function of rich clusters, i.e.~the number density of clusters within a
temperature range $dT$ about $T$.
This needs to be observationally determined over some range of $T$ sufficiently
high that gravitational physics dominates.
Observational and theoretical considerations place this limit at
$T_{\rm X}\,{\simgt}\,3.5\,$keV (e.g.~Finoguenov et al.~2000; 
Nevalainen, Markevitch \& Forman ~2000).
Ideally we could compare the data with a temperature function estimated
directly from a series of large
cosmological hydrodynamic simulations (see e.g.~Pen~1998), which included
all of the physics relevant to determining the emission weighted
inter-galactic medium (IGM) temperature of clusters.
This however is currently computationally infeasible.
Instead we shall make use of the fact that rich clusters are large virialized
structures dominated by gravity, and thus we can factorize the problem and
proceed in two steps.

First we shall use the abundance of dark matter haloes of a given mass drawn
{}from extremely large N-body simulations.
Any dark matter halo large enough to host a cluster is unambiguously seen
in such simulations.
Next we shall use a mass-temperature relation, calibrated from hydrodynamical
simulations, to convert from the (unobservable) virial mass to the IGM
temperature, effectively using the larger volume of the N-body simulations
to improve the statistics of the hydro simulations.
Unfortunately the mass functions determined from the N-body simulations
are somewhat dependent on the method used to define haloes and their masses,
and the precise definition of mass used there
is not exactly what is used in the hydrodynamic
simulations which calibrate the $M{-}T$ relation;
the difference is expected to be small however \cite{JFWCCEY}.
We discuss further details of the $M{-}T$ relation in \S\S\ref{sec:halomass}
and \ref{sec:MT}.

\subsection{The mass variance}

Our constraint will be on the variance of the density field, smoothed on
some (comoving) scale $R$ corresponding to a mass
$M=(4\pi/3)\bar{\rho}R^3$ where $\bar{\rho}$ is the background density.
In terms of the power spectrum
\begin{equation}
  \si^2(R,z)=\int_0^\infty {dk\over k}\ \Delta^2(k,z) W^2(kR)\,,
\label{eq:sir}
\end{equation}
where $\Delta^2 = k^3 P(k,z)/(2\pi^2)$, $P(k)\equiv |\delta_k|^2$ is the
matter power spectrum and $W(kR)$ is the window function corresponding to
the smoothing of the density field (see e.g.~Peebles 1993).
Our mass functions are fitted assuming
a spherical top-hat smoothing, so
\begin{equation}
  W(kR)=\left[ {3j_1(kR)\over kR} \right]\,,
\end{equation}
where $j_1(x)$ is the spherical Bessel function of order 1.
We are interested in both the normalization of the power spectrum, for
which we shall use $\si_8\equiv\sigma(8\,h^{-1}\,{\rm Mpc})$, and its shape.
We use the Cold Dark Matter family of power spectra and parameterize the
shape by $\Gamma$ in the fitting formula of Bardeen et al.~(1986).
While the form of Eisenstein \& Hu~(1999) provides a slightly
better fit to the shape, we will mostly
quote the results in a $\Gamma$ independent way,
rendering this distinction unimportant.

We write $\si(R,z) = \si(R,0)\,g(z)/g(0)$ where the growth factor $g(z)$ can be
computed numerically
\cite{He77,CaPrTu92,Coh,Ham}:
\begin{equation}
  g(z)={5\over2} {\omm \over a} {{d a} \over {d{\tau}}} \int_0^a da^\prime 
  \left({da^\prime} \over {d {\tau}} \right)^{-3},
\label{eq:gz}
\end{equation}
where the scale factor $a=(1+z)^{-1}$, and the dimensionless time
$\tau\equiv H_0 t$.  The Friedmann equation gives
\begin{equation}
  \left({\dot{a}\over a}\right)^2 =  H_0^2\left(
  {\omm\over a^3} + \oml + {\Omega_{\rm K}\over a^2} \right)\,,
\label{eq:Friedmann}
\end{equation}
where $H_0\equiv({\dot a}/a)_{t_0}$ is the Hubble constant.
The usual symbols $\omm$,
$\oml\equiv \Lambda/3H_0^2$ and $\Omega_{\rm K}\equiv 1-\omm-\oml$ are the
density parameters ($\rho/\rho_{\rm crit}$, with
$\rho_{\rm crit}=3H_0^2/8\pi G$)
in matter, cosmological constant and curvature, respectively.
Photons and other relativistic species can be safely ignored.

\subsection{Press-Schechter and modifications}

\begin{figure}
\begin{center}
\epsfig{file=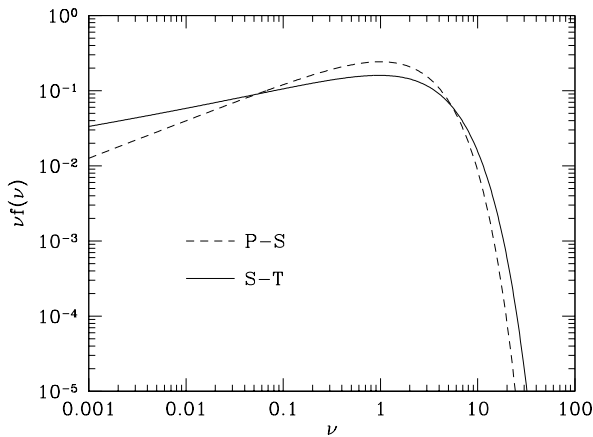, width=8cm}
\end{center}
\caption{A comparison of the Press-Schechter form for the multiplicity
function, used in all previous work, with the Sheth \& Tormen (1999) model,
which better fits the Virgo simulation
results (see text).  Note that Press-Schechter systematically underestimates
the abundance of the most massive cluster haloes, $\nu\sim 10$, and can
either overestimate or underestimate the abundance of more typical clusters
with $\nu\sim$ few.}
\label{fig:fnu}
\end{figure}

Within the PS theory and its extensions the (comoving) number density of
objects of (virial) mass $M$ is a function only of $\sigma(M)$.
Using the scaled variable
\begin{equation}
  \nu \equiv \left( {\delta_{\rm c}\over \sigma(M)} \right)^2\, ,
\end{equation}
with $\delta_{\rm c}\equiv 1.686$, we write the mass function in terms of the
`multiplicity function' $f(\nu)$ as
\begin{equation}
  \nu f(\nu) = {M^2\over \bar{\rho}} {dn\over dM} {d\ln M\over d\ln\nu}
 \,.
\label{eq:PShec}
\end{equation}
In the spirit of using PS as a fitting function to N-body simulations we
do not imbue $\delta_{\rm c}$ with a cosmology dependence (see next section)
but rather keep it constant.  The Press-Schechter model for $f(\nu)$ is
\begin{equation}
  \nu f(\nu) = \sqrt{\nu\over 2\pi} e^{-\nu/2},
\label{eq:fPS}
\end{equation}
and this formula has been extensively used to make predictions of cluster
abundances in different cosmologies.
Recently Sheth \& Tormen~(1999) have proposed a correction to the
Press-Schechter formula, motivated by a model of non-spherical collapse,
that better fits large N-body simulations:
\begin{equation}
  \nu f(\nu) \propto (1+{\widetilde \nu}^{-p})\, {\widetilde \nu}^{\,1/2}
   e^{-{\tilde \nu}/2}\,,
\label{eqn:fnu}
\end{equation}
where $p=0.3$ and ${\widetilde \nu}=0.707\nu$.
The normalization is fixed by the requirement that all of the mass
be in haloes, i.e.
\begin{equation}
  \int f(\nu) d\nu = 1 \,;
\end{equation}
for the above choice of parameters the normalization factor is 0.2162.
Jenkins et al.~\shortcite{JFWCCEY} have shown that Eq.~(\ref{eqn:fnu})
is in very good agreement with the simulations of the Virgo Consortium,
except for very rare objects which shall not be of interest in this work.
We compare the multiplicity functions in Fig.~\ref{fig:fnu}.  Note that
Press-Schechter systematically overestimates the abundance of objects of
cluster mass having
${\rm 1}\,{\simgt}\,\nu\,{\simgt}\,{\rm 6}$ ($\si(R) \simlt 0.7$) 
and underestimates the abundance of objects corresponding to a higher $\si(R)$.

\subsection{Spherical top-hats}

The spherical top-hat ansatz \cite{PPC,LidLyt,Peacock}
models the formation of an object by the evolution of a spherical overdense
region embedded in a homogeneous `background' of mean density $\bar{\rho}$.
This region begins by expanding at the same rate as the background, but since
it is positively curved the expansion slows, comes to a halt and the region
collapses.  Mathematically the evolution proceeds to a point of zero radius,
however physically we assume that virialization occurs at twice\footnote{In
the presence of a cosmological constant there is a small modification to this.}
the turn-around time, resulting in a sphere of half the turn-around radius.
The overdensity (relative to the background) at turn-around is $9\pi^2/16$ for
an Einstein-de Sitter model.
At virialization the background has become less dense and the sphere's density
grown by a further factor of 8 -- we then denote the overdensity relative
to the critical density by $\Delta_{\rm c}$.
This parameter has the value $18\pi^2$ in an Einstein-de Sitter model, and
will in general be a function of $\Omega_{\rm M}$ and $\Omega_\Lambda$.
The extrapolation from linear theory of this overdensity is normally denoted
$\delta_{\rm c}$.
It is $(3/20)(12\pi)^{2/3}\simeq1.686$ for Einstein-de Sitter and varies by a
few percent in other cosmologies.
This density is used as a threshold in PS theory and its extensions.
We shall neglect the small cosmology dependence and simply take
$\delta_{\rm c}$ fixed throughout.

The value of the density contrast at collapse, $\Delta_{\rm c}$, on
the other hand, should be calculated for each model, since it will be
important for the $M{-}T$ relation.
We computed $\Dc$ by numerically integrating the equations of motion for
the spherical top-hat collapse, including the correction to the virial
theorem from the $\Lambda r^2$ potential (Lahav et al.~1991; \S4.2).
Our results can be fit to 2 per cent over the range $0.2\leq\omm\leq 1.1$ and
$0 \leq \oml \leq 1$ by
\begin{equation}
  \Dc=\omm\sum_{i,j=0}^4 c_{ij}\ x^i y^j\,,
\label{eqn:Dcfit}
\end{equation}
where $x\,{\equiv}\,\omm-0.2$, $y\,{\equiv}\,\oml$ and the coefficients
$c_{ij}$ are reported in Table~\ref{tab:coef}.
As an example we plot $\Delta_{\rm c}$ vs $\Omega_{\rm M}$ for flat models,
where the dashed line shows our fit and the solid line is the exact relation.
We also checked this fit with the one provided by Eke, Navarro \& Frenk.~(1998b)
and found an agreement at the 1 per cent level.
Note that some authors use a different convention in which $\Dc$ is specified
relative to the background matter density -- our $\Dc$ is $\omm$ times theirs.

\begin{table}
\caption{\footnotesize%
Coefficients $c_{ij}$ of the fitting formula,
Eq.~(\protect\ref{eqn:Dcfit}), for the collapse overdensity $\Dc$.}
\label{tab:coef}
\begin{center}
\begin{tabular}{l|rrrrr}
\hline
& \multicolumn{5}{|c}{$j$} \\
$i$ & 0 & 1 & 2 & 3 & 4\ \\ \hline
0 & $546.67$  & $-137.82$  &    $94.0830$   &  $-204.680$   &   $111.51$\\
1 & $-1745.6$ & $    627.22$ &    $-1175.2$  &  $  2445.7$ &$     -1341.7$\\
2 & $ 3928.8$ & $   -1519.3$ &    $ 4015.80$  & $  -8415.3$&$      4642.1$\\
3 & $-4384.8$ & $    1748.7$ &    $-5362.1$  &  $  11257.$&$     -6218.2$\\
4 & $ 1842.3$ & $   -765.53$ &    $ 2507.7$  &  $ -5210.7$&$      2867.5$\\
\hline
\end{tabular}
\end{center}
\end{table}

\begin{figure}
\begin{center}
\epsfig{file=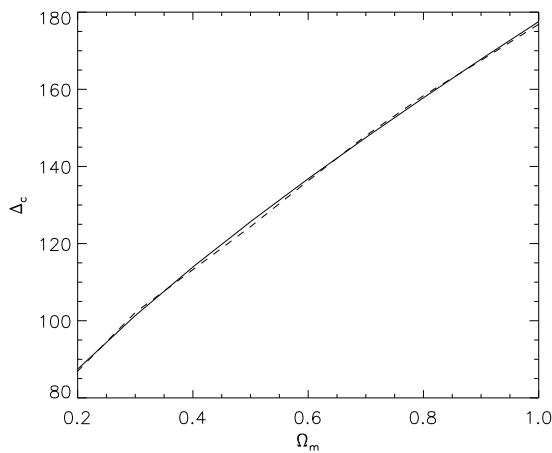 ,width=8cm}
\end{center}
\caption{Showing our fit to the overdensity at virialization, $\Delta_{\rm c}$
vs $\Omega_{\rm M}$ for flat models, where the dashed curve is our fitting
function and the solid curve is the exact result.}
\label{fig:Dc}
\end{figure}

\subsection{Halo mass definitions} \label{sec:halomass}

Before we turn to the mass-temperature relation we note a few important
details about determining the mass of a dark-matter halo.  Unfortunately there
is no unique algorithmic definition of a dark matter halo, even within a 3D
simulation itself.  Some particular group finders are in common use, but no
single group finder is always used.  For large objects such as clusters all
group finders should be able to find all of the clusters, so this is not of
immediate concern, although the degree of substructure will be highly variable
between group finders.

More disconcertingly there are a wide number of definitions of halo
{\it mass\/} in the literature, and they can differ by a large amount.
Even for the mass-temperature relations calibrated by hydro simulations
different authors use
different definitions of `mass', with the differences dependent on the
cosmological parameters.  In this sub-section we briefly review the relevant
mass definitions and discuss which is the most appropriate for
the mass function of Eq.~(\ref{eqn:fnu}).

Although other halo finders are in common use, we shall deal exclusively with
haloes found using the Friends-of-Friends (Davis et al.~1985) algorithm,
hereafter called FOF.  The FOF algorithm has one free parameter, $b$, the
linking length in units of the mean inter-particle spacing.
Commonly used values of $b$ are 0.1, 0.15 and 0.2.
The mass of the halo is simply the sum of the masses of the particles
identified as part of the halo.
An alternative (and more easily interpreted) procedure is to use FOF to find
candidate haloes, identify a halo centre (e.g.~the centre of mass of the halo
or, more robustly, the position of the most bound particle) and then to
calculate the mass from the spherically averaged density profile about that
centre.  This is the technique typically used to define the mass in
hydrodynamic simulations which calibrate the $M\,{-}\,T$ relation.

In this spirit we define $M_\Delta$ as the mass contained within a radius
$r_\Delta$, inside of which the mean interior density is $\Delta$ times the
{\it critical\/} density:
\begin{equation}
  \int_0^{r_\Delta} r^2 dr\ \rho(r) =
  {\Delta\over 3} \rho_{\rm crit} r_\Delta^3
  \,.
\end{equation}
The `virial mass' from the spherical top-hat collapse model would then be
simply $M_{\Delta_{\rm c}}$.
Other masses in common use are $M_{500}$ and $M_{200}$ where the latter is
approximately the virial mass if $\Omega_{\rm M}=1$.

\begin{figure}
\begin{center}
\epsfig{file=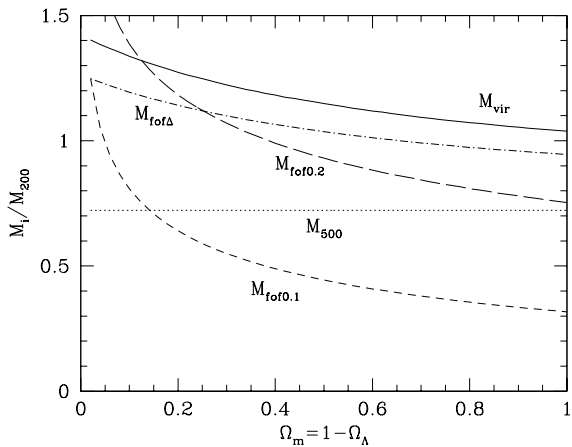 ,width=8cm}
\end{center}
\caption{Relations between various definitions of the mass of a halo as
a function of $\Omega_{\rm M}$ assuming the halo density profile follows the
NFW form with concentration parameter $c=5$.  The FOF based masses are only
crude approximations in this model (see text).}
\label{fig:mdef}
\end{figure}

For large mass haloes many of these definitions are related on average simply
by a factor.  We can estimate this factor by assuming that haloes have a
universal profile, for example the NFW form (Navarro, Frenk \& White~1996):
\begin{equation}
  \rho(r) \propto x^{-1}\left(1+x\right)^{-2},
\end{equation}
where $x=r/r_{\rm s}$ and $r_{\rm s}$ is a scale radius usually specified
in terms of the concentration parameter $c\equiv r_{200}/r_{\rm s}$.
Navarro et al.~(1996) refer to $r_{200}$ and $M_{200}$ throughout as
the `virial radius' and `virial mass' respectively.  Again, N-body simulations
have shown that the concentration parameter is a weak function of virial mass,
having the value $c\sim 5$ for masses characteristic of clusters.
We can use this profile to relate the various mass definitions, as shown in
Fig.~\ref{fig:mdef}.

Unfortunately, while this model works well for converting between spherically
averaged mass definitions based on $M_{\Delta}$, there is a large scatter for
masses based on group membership, such as $M_{{\rm fof}0.2}$.  This is
because with $b=0.2$, FOF can link together neighbouring haloes in
supercluster-like structures, increasing the mass assigned to the structure
compared to the $M_\Delta$ estimators.
Thus the `FOF' lines in Fig.~\ref{fig:mdef} must be taken as highly uncertain.
Comparing the mass function from a high-resolution N-body simulation (of the
Ostriker \& Steinhardt~1995 concordance model) with the universal
form of Jenkins et al.~\shortcite{JFWCCEY} we find the best match is obtained
if we interpret their mass as $M_{\Delta_{\rm c}}$.
We shall assume below that this remains true independent of cosmology.
But we note that this ambiguity in the definition of mass remains a
significant source of uncertainty.

\subsection{The M--T relation} \label{sec:MT}

The mass function of rich clusters is itself not observable.
However the local temperature function is reasonably well known
(see \S\ref{sec:data}).  To predict the latter from the former we need a
relation between emission weighted IGM temperature and cluster virial mass.
Our results are quite sensitive to the choice of $M{-}T$ relation, and
currently the uncertainty in this relation is the largest theoretical source
of error in determining $\sigma_8$ (see also Voit~2000).

Recent observational determinations of the mass-temperature relation
\cite{HorMS,NevMF} disagree at the several tens of percent level
(in mass at fixed temperature) when using different estimators of the cluster
mass.  The virial mass of a cluster is a notoriously difficult quantity to
obtain observationally with high accuracy -- while different estimators
clearly correlate well, they disagree at the level of accuracy required here.
Such observations do however provide general support for the functional form
and scalings predicted by the spherical collapse model (see below) for
clusters of sufficiently large mass.
Similar scalings are seen in hydrodynamic simulations of galaxy clusters in
a cosmological context.  In the simulations the total mass of a cluster is
easy to obtain (though convention dependent, \S\ref{sec:halomass}) and in what
follows we shall use a $M{-}T$ relation derived from simulations.
While these hydrodynamic simulations show good agreement for the total mass
and X-ray temperature properties of clusters (Frenk et al.~1999)
there are several uncertainties which enter when comparing the simulations to
observations and which are important to note.

As with all simulation derived results there are issues related to numerical
convergence.  With the latest round of high resolution simulations the
situation in this respect has improved dramatically.  However, the spectrally
measured cluster temperatures may not coincide with the mass or emission
weighted temperature estimated from the simulations, due to the influence of
soft line emission \cite{MatEvr}.
Secondly, most simulations are done with purely adiabatic hydrodynamics, which
ceases to be a good model for the lower mass/temperature clusters.
It is also possible that energy injection (possibly from SNe, AGN or galaxies)
has altered the $M{-}T$ relation, again an effect thought to operate
preferentially on lower mass/temperature clusters.
Thirdly the simulations usually predict the emission weighted temperature,
which only converges if data are used out to a radius ${\sim}\,r_{500}$.
Observers often probe different ranges of radius in determining the cluster
temperature and perform more sophisticated modelling, which can lead to
discrepancies between the measured and simulated temperatures.
Finally, there still remain significant (for our purposes) calibration
uncertainties for the detectors.

With these caveats in mind, the results from the simulations can be quoted
in terms of corrections to the spherical collapse model which relates the mass
to the (virial) temperature of the hot IGM.
For an object virialized at a redshift $z$ we have\footnote{Our definition
of $\beta$ differs from that of Henry~\protect\shortcite{Henry}:
$\beta_{\rm Henry}=1.42/\beta$, and our $\beta$ should not be confused with
the slope of the emissivity profile of clusters.}
\begin{eqnarray}
  \left( {M(T,z)\over 10^{15}\,h^{-1}\,{\rm M}_\odot }\right)
&=& \left( {T\over\beta} \right)^{3/2}
    \left(\Dc E^2 \right)^{-1/2} \times \nonumber \\
& & \left[ 1 -  2  {\oml(z)\over\Dc  }\right]^{-3/2}\ ,
\label{eqn:m-t}
\end{eqnarray}
where $T$ is in keV, $\Dc$ is the mean overdensity inside the virial radius
in units of the critical density and, from Eq.~(\ref{eq:Friedmann}),
$E^2=\omm (1+z)^3 + \oml + \Omega_{\rm k} (1+z)^2$.
Note that $\Dc$ is a redshift dependent variable, and should be evaluated
using the appropriate $\oml(z)$ and $\omm(z)$.
The term in square brackets is a correction to the virial relation arising
{}from the additional $r^2$ potential in the presence of $\Lambda$
\cite{LRLP,VL96,WanSte}.
It provides only a small correction and, though we include it, it can be
neglected at the present level of accuracy.

\begin{table}
\caption{\footnotesize%
The mass-temperature relation determined from hydrodynamical simulations.
The quoted value of $\beta$ is that relevant for use in
Eq.~(\protect\ref{eqn:m-t}).  From top to bottom the
references are: EMN \protect\cite{EMN}; ENF \protect\cite{ENF};
BN \protect\cite{BryNor}; YJS \protect\cite{YosJS};
TC \protect\cite{TC} and Tetal \protect\cite{Thometal}.
We have quoted emission weighted temperatures where available, while the last
two values (marked with an asterisk) are core temperatures.}
\label{tab:hydro-mt}
\begin{center}
\begin{tabular}{l|ccccc}
\hline
Name & $\Omega_{\rm M}$ & $\Omega_\Lambda$ & $\Omega_{\rm B}$ & h & $\beta$ \\
\hline
\vspace{2pt}
EMN        & 1.0 & 0.0 & 0.10 & 0.50 & 1.21 \\
EMN        & 0.2 & --- & 0.10 & 0.50 & 1.42 \\
ENF        & 0.3 & 0.7 & 0.04 & 0.70 & 1.33 \\
BN         & 1.0 & 0.0 & 0.06 & 0.50 & 1.10 \\
BN         & 1.0 & 0.0 & 0.10 & 0.65 & 1.04 \\
BN         & 1.0 & 0.0 & 0.08 & 0.50 & 1.04 \\
BN         & 0.4 & 0.0 & 0.06 & 0.65 & 1.08 \\
YJS        & 0.3 & 0.7 & 0.03 & 0.70 & 1.48 \\
TC$^{*}$   & 1.0 & 0.0 & 0.10 & 0.65 & 1.61 \\
Tetal$^{*}$& 1.0 & 0.0 & 0.06 & ---  & 1.23 \\
\hline
\end{tabular}
\end{center}
\end{table}

Ideally the normalization and scatter of the $M{-}T$ relation would be
determined by simulations, so that theoretical models can be compared with
the data rather more directly (e.g.~Pen~1998).
However Eq.~(\ref{eqn:m-t}) is a remarkably good fit to the simulations, which
are sufficiently computationally demanding that they cannot explore parameter
space efficiently.  Thus we rely on a hybrid approach where the coefficients
are determined from simulations, while the scalings are taken from simple
theoretical models \cite{Mat00}.
Specifically we use the hydrodynamic simulations to determine $\beta$,
together with Eq.~(\ref{eqn:m-t}) for the mass and redshift dependence.
In practice we used the $M{-}T$ relation at an observed $z=0.053$,
corresponding to the median redshift of our cluster sample
(see \S\ref{sec:data}).  We then shift the resulting $\sigma(z=0.053)$ value 
to $z=0$ using the growth rate, Eq.~(\ref{eq:gz}), which is a significant
5 per cent correction.

The simulations normalize $\beta$ assuming that the virialization redshift
is the redshift of observation.  In principle one could attempt to correct
for the virialization redshift dependence, however we have chosen not to do
this, and here we differ from Viana \& Liddle \shortcite{VL98} and
Wang \& Steinhardt (1998), for example.
The simulations, which clearly include the full effects of variations
in the virialization redshift, give an $M{-}T$ relation well fit by
Eq.~(\ref{eqn:m-t}) if $z$ is interpreted as the redshift of observation.
We believe that 
the effect of differing virialization redshifts {\it is included in the
simulations as part of the scatter\/}
about the mean relation (see below), and so to add an
additional effect by hand would be incorrect.
Comparison of the scatter in the $M{-}T$ relation found by
Bryan \& Norman~(1998) in a full cosmological simulation with that
of Evrard, Metzler \& Navarro~(1996), who use constrained realizations
(and thus constrained formation times), suggests that in fact
the effect of scatter in the virialization redshift is a very
small source of the total scatter in the relation.
Most of the scatter is due to the different merging histories
(see also Cavaliere, Menci \& Tozzi~1999).
While further simulations will be needed to address this issue properly,
recent work (Mathiesen~2000) suggests that minor merger
events have a more important influence on the evolution of the temperatures
than major mergers (and hence formation time).

A summary of recent numerical experiments which constrain $\beta$ is given
in Henry~\shortcite{Henry}.
We have taken Henry's list, added some recent work and
corrected one of the relations to use $M_{\rm vir}$.
Our results are shown in Table~\ref{tab:hydro-mt}.
The $M{-}T$ relation of Evrard et al.~\shortcite{EMN}
defines mass as $M_{200}$, so that in the
context of the spherical top-hat model there is predicted to be an
$\Omega_{\rm M}$ dependence to the prefactor of the scaling relation.
Correcting for this scaling brings the $\beta$s obtained in these simulations
into better agreement, but they still disagree slightly, suggesting that the
scaling is only approximately observed.
We postulate that this is because $\Omega_{\rm B}/\Omega_{\rm M}$ changes
drastically between the simulations.
The $M$s of Eke et al.~\shortcite{ENF}, Bryan \& Norman \shortcite{BryNor},
and Yoshikawa et al.~\shortcite{YosJS}, on the other hand, are already the
`virial' mass in the sense of the spherical top-hat model.
Also quoted in Table~\ref{tab:hydro-mt} are the $\beta$ values from
Tittley \& Couchman \shortcite{TC} and Thomas et al.~\shortcite{Thometal}.
While all of the other temperatures in the Table are
emission weighted, these authors use the average temperature within a core
region, which could be slightly different.

As shown in Table~\ref{tab:hydro-mt} the values of $\beta$ spread from
near $1$ up to $1.6$ with no obvious peak of preferred values.
We adopt the mean opinion on this issue, allowing for $\beta = 1.3$, with 
a 10 per cent systematic error (i.e.~variation between the simulations)
in the mass, plus a 10 per cent statistical error.
The second error arises from the intrinsic scatter in individually determined
$M{-}T$ relations, and is mostly due to the merging history of the
clusters.  Most of the simulations agree on the scatter about the mean relation
quite well, though Eke et al.~\shortcite{ENF}
find a slightly enhanced scatter compared to the other authors.
This dispersion is accurately modelled as a Gaussian, and we treat the
systematic uncertainty as a Gaussian also.  The $\beta$ value advocated
by Henry~\shortcite{Henry} is slightly lower, $\beta=1.17$,
with a suggested systematic error of 4.1 per cent, while
the value we would obtain using the observational determinations is close to
unity.

\subsection{Summary of modelling}

In summary, we use the mass function of Jenkins et al.~\shortcite{JFWCCEY},
interpretting the
mass as the top-hat virial mass.
For an object at fixed mass we randomly assign a temperature using
Eq.~(\ref{eqn:m-t}) where $\Dc$ is given by Eq.~(\ref{eqn:Dcfit}) and 
the $M{-}T$ relation is evaluated at a median redshift
interpreted as the redshift of observation.
{}From Table~\ref{tab:hydro-mt} we choose $\beta=1.3\pm 0.13\pm 0.13$  with
Gaussian errors.
Here the first error represents the scatter about the mean relation and
the second is the systematic uncertainty in the prefactor from the different
calculations.  We carry out the comparison between theory and data at
the median redshift of the observations, and then correct the final
normalization to the appropriate $z=0$ value.

\section{Data} \label{sec:data}

\subsection{Definition of cluster sample}

There has been progress recently in observationally determining the
temperature function of nearby clusters, and we have compiled from the
literature a new sample with which to constrain the normalization of
the power spectrum at $z\,{\simeq}\,0$.

The sample is adapted from the 30 clusters compiled by
Markevitch~\shortcite{Mar}.
His sample was selected from bright clusters having flux above
$2\times10^{-14}{\rm W}\,{\rm m}^{-2}$ in the \ROSAT\ 0.2--2\,keV band, and
over the redshift range $z=0.04$--0.09.
The flux limit is a factor of 4 above the nominal flux limit of 
the \ROSAT\ Brightest Cluster Sample \cite{RBCS},
and hence the sample is expected to be close to complete for these fluxes.
The upper redshift limit is imposed by the lack of clusters bright enough to
be detected, and the lower limit by \ROSAT\ selection effects.
Clusters are also excluded with galactic latitude $|b|<20^\circ$, where
observations become affected by the Galaxy.  This sample is close to
volume limited at the high-temperature end, and the numbers can be corrected
for incompleteness at the low-temperature end by using the
effective volume (see \S\ref{sec:likelihood}).
Markevitch~\shortcite{Mar} determined temperatures by excising the
central regions from clusters to approximately correct for the effects of
cooling flows, and used a hybrid approach of combining \ROSAT\ emissivity
profiles with \ASCA\ data to fit $T_{\rm X}$.
As we discuss below, we believe that somewhat better temperatures are now
available based on careful fitting of \ASCA\ data alone.
Using mainly these temperatures, together with a somewhat different selection,
we have attempted to define an {\it effectively temperature selected\/}
sample.

White \& Buote~\shortcite{WhiBuo} use a maximum likelihood method to determine
radial temperature profiles for clusters using \ASCA\ data alone.
This includes a Monte Carlo method for redistribution of X-ray photons by
\ASCA's complex optics.
They account for cooling flows using a single extra parameter in their fits,
and find that the cooling flow corrected temperatures are generally consistent
with what would be fitted to the outer regions of the clusters, although with
less uncertainty.
They also do not find the temperature fall-off at large radii found by
Markevitch~\shortcite{Mar}, about which there has been much discussion
in the literature (e.g.~Irwin, Bregman \& Evrard~1999).  We regard
the temperatures presented in White~\shortcite{WhiteD} as representing
the most careful analysis of X-ray temperatures from available \ASCA\ data.
Independent \BeppoSAX\ data, available for about a quarter of these clusters
give temperatures in good agreement \cite{IrwBre}.
These values are unlikely to improve significantly until \Chandra\ and
\XMM\ data become widely available.

We have constructed our cluster sample in a similar way to that given by
Markevitch \shortcite{Mar},
although we use temperatures from White \shortcite{WhiteD} when available,
supplemented by other temperatures from the literature.
The sample of course has a large overlap with other low-redshift cluster
samples, such as those of Edge et al.~\shortcite{Edgetal}, Henry \& Arnaud
\shortcite{HA91}, Henry \shortcite{Henry}
and Blanchard et al.~\shortcite{BlaSBL} -- although we
neglect the lowest redshift clusters for reasons of incompleteness,
as well as concerns about biases introduced by sample variance.

Since many of the White~\shortcite{WhiteD} temperatures show significant
changes compared with Markevitch \shortcite{Mar}, we also need to reconsider
the completeness of the sample.
We use the estimated $L_{\rm X}{-}T_{\rm X}$ relationship derived from
Markevitch \shortcite{Mar} to decide whether a cluster would be above the
\ROSAT\ flux limit based on the improved value of $T_{\rm X}$.
Since we find that the White \shortcite{WhiteD} temperatures are a factor
${\simeq}\,1.14$ higher than the Markevitch \shortcite{Mar} temperatures on
average for the clusters in common, we correct the $L_{\rm X}{-}T_{\rm X}$
relation by this factor.  Explicitly we use
\begin{equation}
  L_{\rm X}=1.07\times10^{37}\left({T_{\rm X}\over 6\,{\rm keV}}\right)^{2.1}
  h^{-2}{\rm W},
\label{eqn:lumtemp}
\end{equation}
where as usual $h=H_0/(100\,{\rm km}\,{\rm s}^{-1}{\rm Mpc}^{-1})$.
Markevitch~\shortcite{Mar} finds a relatively small scatter about this
relationship once cooling-flow effects have been corrected for.
(A steeper temperature dependence is found for bolometric X-ray
luminosity or luminosity over higher energy ranges.)
The use of this relationship allows us to avoid using any specific
{\it flux\/} information for a particular cluster, which is advantageous since
this information is somewhat uncertain, varying significantly between
instruments and between analysis methods.
Provided Eq.~(\ref{eqn:lumtemp}) is approximately correct, the details will
be a higher order correction to our $\sigma_8$ constraints.

\begin{figure}
\begin{center}
\epsfig{file=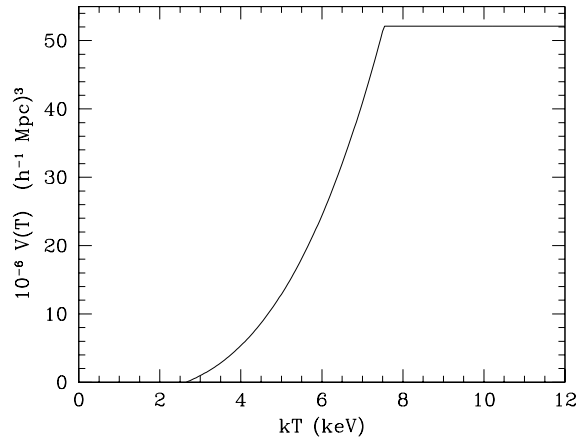 ,width=8cm}
\end{center}
\caption{The effective volume of the sample as a function of temperature.
Note that the sample is volume limited for clusters above about $7.5\,$keV,
and that the correction factor is not too drastic for $kT\simgt3.5\,$keV.}
\label{fig:vT}
\end{figure}

Our sample is thus effectively {\it temperature\/}-selected at each redshift,
and hence we will only need to correct for the volume sampled at each
temperature over the full redshift interval.
We use CMB-frame redshifts from the recent compilation of
Struble \& Rood~\shortcite{StrRoo} for the Abell clusters, and the redshifts
given in White \shortcite{WhiteD} or Ebeling et al.~\shortcite{RBCS} otherwise.
At these low redshifts only a small error is made by assuming non-expanding
Euclidean space, since the correction to the luminosity distance is
${\cal O}(z/4)$, which is ${\sim}\,2$ per cent at worst.
It is easy to compute the relationship exactly, although this would in
principle have to be re-calculated for each model.
We use redshift as an exact distance indicator, which is another reason to
cut off the sample at low redshifts, where this will cease to be a good
approximation.  The effective volume, as a function of temperature, is shown
in Fig.~\ref{fig:vT}.  Because of the large completeness correction required
at low temperature, we make a further cut of all clusters with
$T_{\rm X}<3.5\,$keV.
This also corresponds approximately to a restriction to clusters with
temperatures which are dominated by gravitational physics
\cite{MohEvr,BalBP,XueWu}.
Our final sample of 38 clusters is presented in Table~\ref{tab:data}.

\subsection{Supplementary sample}

Since we will be considering the temperature errors in our Monte Carlo
error analysis (\S\ref{sec:MonteCarlo}), some clusters will scatter
out of the selection cuts and hence
it is important to include clusters which could
scatter {\it into\/} the sample.  We include in Table~\ref{tab:extradata}
additional clusters with $T_{\rm X}+2\sigma$ being above our 3.5\,keV
cut-off, or above the temperature-derived flux limit at their redshift.
This limit corresponds to
\begin{equation}
{T_{\rm X}\over{\rm keV}}>2.65 \left({z\over0.03}\right)^{0.95},
\end{equation}
and is indicated in Fig.~\ref{fig:selection} by the roughly diagonal line.
To populate this supplementary sample (as well as to check completeness of
the main sample) we extended the \ROSAT\ selection down a further factor of
2 to $1.0\times10^{-14}{\rm W}\,{\rm m}^{-2}$.
In particular we scrutinized the BCS sample of
Ebeling et al.~\shortcite{RBCS}, together with the RASS1 Bright Cluster
Survey of de Grandi et al.~\shortcite{RSouth} for the southern extension,
and the overlapping XBACS sample \cite{XBACS}
came from either White~\shortcite{WhiteD}
or the David et al.~\shortcite{David} compilation when available.
The REFLEX survey (see B{\" o}hringer et al.~2000)
may ultimately be better for this purpose, but is not yet published.
For some clusters we used temperatures from the 
White, Jones \& Forman~\shortcite{WJF} deprojection modelling of \Einstein\
data.  For a few cases we had to resort to estimates \cite{RBCS}
based on the X-ray luminosity -- for those cases we adopted a representative
error of $\pm1.00\,$keV, which flags them in Table~\ref{tab:extradata}.

\subsection{Comparison with other samples}

\begin{figure}
\begin{center}
\epsfig{file=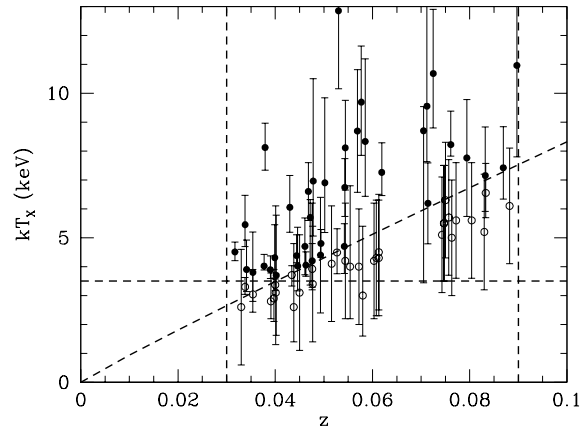 ,width=8cm}
\end{center}
\caption{An indication of our selection function (dashed lines).
We chose clusters (solid
symbols) with CMB-frame redshifts within $0.03\,{<}\,z\,{<}\,0.09$ and a
temperature-dependent function to account for a flux limit.  We also
cut out all clusters below $kT_{\rm X}\,{=}\,3.5\,$keV.  {\it However},
we kept clusters (shown as open circles) which could scatter into our
acceptance region within their $\pm2\sigma$ temperature uncertainties
(which are plotted).
Redshift errors are here assumed to be negligible.  The lack of clusters
at $z\simeq0.065$ is a well-known effect caused by a genuine lack of
superclusters at those distances.}
\label{fig:selection}
\end{figure}

Our final sample differs from that of Markevitch~\shortcite{Mar} in several
details.  We chose to widen the sample down to $z=0.03$, since that increased
the statistics, while we could find no evidence that there was any significant
bias introduced.
This can be seen in Fig.~\ref{fig:selection}, where we have shown the clusters
(solid points) together with our selection cuts in redshift and in temperature.
Clusters which could scatter into our sample are indicated by open symbols.

The White~\shortcite{WhiteD} temperatures were generally a little higher than
those of Markevitch \shortcite{Mar}, resulting in several clusters entering
our sample, explicitly
Abell\,193, Abell\,376, Abell\,1775, Abell\,2255,
Abell\,3532 and IIZw108.
On the other hand Abell\,780, Abell\,1650 and MKW3s
were too cool at their redshifts, while
Abell\,3112 and 2A\,0336 even failed the $2\sigma$ 3.5\,keV cut.
Abell\,2199, Abell\,2634 and Abell\,4038, which are in some other samples,
were lost here because their CMB-frame redshifts \cite{StrRoo} --
as opposed to the heliocentric redshifts more commonly given --
are slightly less than 0.03, while
Abell\,3921 is higher than 0.09.
Furthermore we added Abell\,496, Abell\,576, Abell\,2063, Abell\,2107,
Abell\,2147, Abell\,2151a,
at $0.03<z<0.04$.

Our final sample of 38 clusters is presented in Table~\ref{tab:data}, where
we list the common name, redshift (taken from Struble \& Rood~1999
or White~2000), X-ray temperature and $1\sigma$ error bar.
These temperatures are mostly from White~\shortcite{WhiteD}, but in a few
cases {}from
Markevitch \shortcite{Mar} errors scaled from 90 per cent confidence,
or from the 80 per cent confidence region estimates of
White et al.~\shortcite{WJF}.
Table~\ref{tab:extradata} lists the additional clusters which could scatter
into our sample within their temperature uncertainties.
Through this selection process we believe we have a reasonably constructed
{\it temperature}-selected sample.

\begin{table}		
\caption{\footnotesize%
Our nearby cluster sample.  The data are adapted from the sample of
Markevitch~\protect\shortcite{Mar} with White~\protect\shortcite{WhiteD}
temperatures, and Struble \& Rood~\protect\shortcite{StrRoo} redshifts,
as discussed in the text.  Reported errors are 1$\sigma$.
Note that the White~(2000) method may overestimate the
temperature for some clusters for which the existence of a central cooling
flow is debatable, e.g.~A399, A401, A754, A1775, A3158 and A3562
(M.~Markevitch, private communication).  Removing these clusters has
negligible effect on our results, and we prefer to use the White~(2000)
temperatures, when available, for the sake of consistency.}
\label{tab:data}
\begin{center}
\begin{tabular}{lcc}
\hline
Name	&    $z$   & $T_X$ \\  \hline
A85	&  0.0543& $ \pho6.74^{+0.50}_{-0.50}$\\
A119	&  0.0430& $ \pho6.05^{+0.55}_{-0.43}$\\
A193    &  0.0476& $ \pho4.20^{+1.00}_{-0.50}$\\
A376    &  0.0472& $ \pho5.70^{+0.31}_{-1.17}$\\
A399	&  0.0712& $ \pho9.55^{+1.92}_{-0.96}$\\
A401	&  0.0725& $10.68^{+1.11}_{-0.94}$\\
A478	&  0.0869& $ \pho7.42^{+0.71}_{-0.54}$\\
A496    &  0.0317& $ \pho4.51^{+0.17}_{-0.15}$\\
A576    &  0.0377& $ \pho4.02^{+0.20}_{-0.07}$\\
A754	&  0.0530& $12.85^{+1.77}_{-1.35}$\\
A1644	&  0.0461& $ \pho4.70^{+0.49}_{-0.49}$\\
A1651	&  0.0832& $ \pho7.15^{+0.84}_{-0.62}$\\
A1736	&  0.0446& $ \pho4.02^{+0.67}_{-0.43}$\\
A1775   &  0.0705& $ \pho8.70^{+0.42}_{-2.63}$\\
A1795	&  0.0619& $ \pho7.26^{+0.51}_{-0.40}$\\
A2029	&  0.0761& $ \pho8.22^{+0.58}_{-0.20}$\\
A2063   &  0.0341& $ \pho3.90^{+0.51}_{-0.38}$\\
A2065	&  0.0714& $ \pho6.19^{+0.70}_{-0.70}$\\
A2107   &  0.0399& $ \pho4.31^{+0.57}_{-0.35}$\\
A2142	&  0.0897& $10.96^{+2.56}_{-1.58}$\\
A2147   &  0.0338& $ \pho5.45^{+0.51}_{-0.38}$\\
A2151a  &  0.0354& $ \pho3.80^{+0.70}_{-0.50}$\\
A2255   &  0.0794& $ \pho7.76^{+1.01}_{-1.01}$\\
A2256	&  0.0569& $ \pho8.69^{+1.06}_{-1.06}$\\
A2589	&  0.0402& $ \pho3.70^{+1.04}_{-1.04}$\\
A2657	&  0.0390& $ \pho3.89^{+0.24}_{-0.15}$\\
A3158	&  0.0585& $ \pho8.33^{+1.43}_{-0.95}$\\
A3266	&  0.0577& $ \pho9.69^{+0.97}_{-0.92}$\\
A3376	&  0.0444& $ \pho4.38^{+0.36}_{-0.12}$\\
A3391	&  0.0502& $ \pho6.90^{+1.47}_{-0.86}$\\
A3395	&  0.0494& $ \pho4.80^{+0.24}_{-0.24}$\\
A3532   &  0.0542& $ \pho4.70^{+0.39}_{-0.47}$\\
A3558	&  0.0468& $ \pho6.60^{+0.50}_{-0.50}$\\
A3562	&  0.0478& $ \pho6.96^{+1.77}_{-0.95}$\\
A3571	&  0.0379& $ \pho8.12^{+0.42}_{-0.39}$\\
A3667	&  0.0544& $ \pho8.11^{+0.82}_{-0.73}$\\
A4059	&  0.0463& $ \pho4.05^{+0.23}_{-0.19}$\\
IIZw108 &  0.0493&  $ \pho4.40^{+1.00}_{-1.00}$\\
\hline
\end{tabular}
\end{center}
\end{table}

\begin{table}		
\caption{\footnotesize%
Additional clusters which could scatter above our selection
criteria within their $\pm2\sigma$ temperature uncertainties.
Additional temperatures come from David~(1993) and White et al.~(1997).
Those with errors given as $\pm1.00\,$keV are estimates from X-ray
luminosities.  }
\label{tab:extradata}
\begin{center}
\begin{tabular}{lcc}
\hline
Name	&    $z$   & $T_X$ \\  \hline
A133    &  0.0554& $ \pho4.00^{+1.40}_{-0.90}$\\
A168    &  0.0438& $ \pho2.60^{+1.10}_{-0.60}$\\
A780	&  0.0527& $ \pho4.49^{+0.41}_{-0.37}$\\
A1650	&  0.0833& $ \pho6.55^{+0.51}_{-0.43}$\\
A1800   &  0.0743& $ \pho5.10^{+1.00}_{-1.00}$\\
A1831   &  0.0603& $ \pho4.20^{+1.00}_{-1.00}$\\
A2052   &  0.0338& $ \pho3.30^{+0.16}_{-0.13}$\\
A2061   &  0.0772& $ \pho5.60^{+1.00}_{-1.00}$\\
A2151   &  0.0354& $ \pho3.04^{+0.43}_{-0.31}$\\
A2249   &  0.0804& $ \pho5.60^{+1.00}_{-1.00}$\\
A2495   &  0.0763& $ \pho5.00^{+1.00}_{-1.00}$\\
A2572a  &  0.0391& $ \pho2.80^{+0.49}_{-0.30}$\\
A2593   &  0.0401& $ \pho3.10^{+1.50}_{-0.90}$\\
A2665   &  0.0544& $ \pho4.20^{+1.00}_{-1.00}$\\
A2734   &  0.0613& $ \pho4.50^{+1.00}_{-1.00}$\\
A3528a  &  0.0516& $ \pho4.10^{+1.00}_{-1.00}$\\
A3560   &  0.0477& $ \pho3.40^{+1.00}_{-1.00}$\\
A3695   &  0.0882& $ \pho6.10^{+1.00}_{-1.00}$\\
A3716   &  0.0450& $ \pho3.10^{+1.00}_{-1.00}$\\
A3809   &  0.0608& $ \pho4.30^{+1.00}_{-1.00}$\\
A3822   &  0.0747& $ \pho5.50^{+1.00}_{-1.00}$\\
A3822   &  0.0747& $ \pho5.50^{+1.00}_{-1.00}$\\
A3880   &  0.0572& $ \pho4.00^{+1.00}_{-1.00}$\\
EXO0422 &  0.0397& $ \pho2.90^{+0.50}_{-0.40}$\\
MKW3s	&  0.0434& $ \pho3.71^{+0.16}_{-0.19}$\\
RXJ1733 &  0.0330& $ \pho2.60^{+1.00}_{-1.00}$\\
S0405   &  0.0613& $ \pho4.30^{+1.00}_{-1.00}$\\
S1101   &  0.0580& $ \pho3.00^{+1.20}_{-0.70}$\\
SC1327  &  0.0476& $ \pho3.92^{+0.44}_{-0.32}$\\
Zw235   &  0.0830& $ \pho5.20^{+1.00}_{-1.00}$\\
Zw5029  &  0.0750& $ \pho6.30^{+1.00}_{-1.00}$\\
Zw8276  &  0.0757& $ \pho5.70^{+1.00}_{-1.00}$\\ 
Zw8852  &  0.0400& $ \pho3.37^{+0.10}_{-0.10}$\\
\hline
\end{tabular}
\end{center}
\end{table}

\begin{figure}
\begin{center}
\epsfig{file=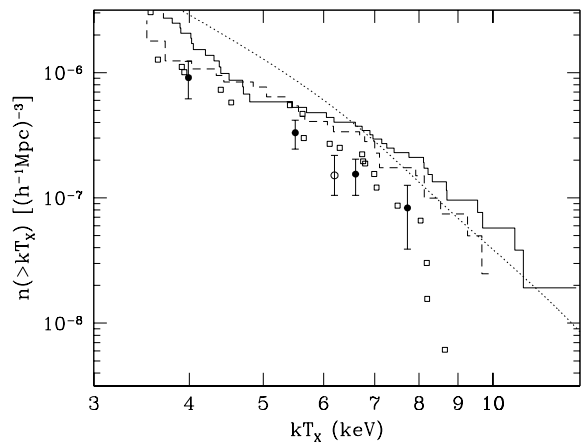 ,width=8cm}
\end{center}
\caption{The temperature function constructed from the sample in
Table~\protect\ref{tab:data} (solid line),
compared to some earlier estimates.
The dashed line is from Markevitch \protect\shortcite{Mar}, the
boxes from Henry~\protect\shortcite{Henry}, the filled circles from
Eke et al.~\protect\shortcite{ECFH},
and the open circle from Viana \& Liddle \protect\shortcite{VL98}.
The curved line is our best fit for a flat cosmology with $\omm = 0.3$}
\label{fig:tempfn}
\end{figure}

Though we do not use it directly in the analysis (see \S\ref{sec:likelihood})
we have constructed the temperature function of our sample for comparison
with earlier work.  We show this in Fig.~\ref{fig:tempfn}, where we have
only used the clusters in Table~\ref{tab:data}.  Some of the other estimates
plotted were derived for specific cosmological models or at other redshifts, so
they cannot be compared in great detail.  However, it is clear that we are in
general agreement with the temperature function estimated by
Markevitch \shortcite{Mar}, although a little higher.  More striking is
that both of these estimates are considerably higher than
those of Henry~\shortcite{Henry} and of Eke et al.~\shortcite{ECF} and Viana
\& Liddle \shortcite{VL98} which are based on the earlier Henry \& Arnaud
\shortcite{HA91}
sample.  The main reason for this difference is the correction
for cooling flows.  We believe that the most physically realistic
comparison of the $z\,{\simeq}\,0$ clusters with the results of simulations
is to correct for cooling flows, since the full cooling-flow physics is
not contained in the simulations.\footnote{The cooling flow correction applied
by White~(2000) may overestimate the temperature for some clusters like A754.
However, we decided to stick to the White~(2000) results in all cases, for
reasons of consistency.}  This gives rise to an added complication
when carrying out evolutionary studies of the high-$z$ vs low-$z$ samples,
since detailed information about cooling flows in high-$z$ clusters tends not
to exist.  However,
for our purposes it seems clear that we should use cluster temperatures
which have been corrected as carefully as possible for the effects of
cooling flows.

\section{Statistical approach} \label{sec:stats}

\subsection{The likelihood function} \label{sec:likelihood}

Rather than fitting to a measure of $n({>}\,kT_{\rm X})$ at some fiducial
$T_{\rm X}$, we try to fit to all the data in a way which accounts for the
Poisson distribution of clusters at each temperature,
using a likelihood method which takes full account of the errors.
Our approach is similar to that
of some other studies (e.g.~Eke et al.~1996; Markevitch~1998;
Henry~2000; Blanchard et al.~2000), but we describe
it here in full, so that the differences can be appreciated.

Given a set of data with which to compare (\S\ref{sec:data}) we compute
the likelihood function for any given theory as follows.  We break the range
of temperatures under consideration into a large number of bins, chosen to be
narrow enough so that the probability of two clusters occupying the same bin
is very small.
Then for a given Monte-Carlo realization (\S\ref{sec:MonteCarlo}) of the
temperatures of the set of clusters we place the clusters into the appropriate
temperature bins, giving an occupation number $\eta_i=0$ or 1 for each bin.
For a predicted number density ${\cal N}_i=n(T)dT$, depending on our
cosmological parameters, the mean occupation number of each bin is
$\mu_i={\cal N}_i V_i \ll 1$.  Here $V_i$ is the volume of space
to which clusters in bin $i$ can be seen:
\begin{equation}
  V_i = {\Omega_{\rm s}\over 3}\left( d_2^3-d_1^3 \right)\,,
\end{equation}
if the survey solid angle is $\Omega_{\rm s}$, which is
$4\pi(1-\cos70^\circ)\simeq8.27\,{\rm sr}$ here.  With a Euclidean
assumption $d_1=cz_1/H_0$, while
\begin{equation}
  d_2 = {\rm min}\left( {cz_2\over H_0},
                        \sqrt{ L(T_{\rm X})\over 4\pi f_{\rm lim} } \right)\,,
\end{equation}
where $f_{\rm lim}$ is the limiting flux of the survey
($2\times10^{14}{\rm W}\,{\rm m}^{-2}$) and $L(T_{\rm X})$ is obtained
{}from the luminosity-temperature relation, Eq.~(\ref{eqn:lumtemp}), given the
cluster's temperature.
Then the likelihood of observing this combination of clusters
is simply
\begin{equation}
\ln{\cal L} = \sum_i \left[(\eta_i-1)\mu_i
 + \eta_i \ln{(1 - \exp(- \mu_i))}\right] \,,
\end{equation}
where by assumption only one of the two terms is non-zero for each bin $i$.
This correctly accounts for the Poisson errors and uses the full temperature
information from the sample.

\subsection{Monte Carlos} \label{sec:MonteCarlo}

Given that there are non-negligible uncertainties occurring
in several places, that the calculation is quite non-linear, and that $n(T)$
is a steeply falling function, it is important to treat errors carefully.
The only reasonable way to do this is through a Monte Carlo approach
(also emphasized by Viana \& Liddle~1999 and
Blanchard et al.~2000), which we now describe.
Firstly, we choose a temperature for each cluster (in Tables~\ref{tab:data}
and \ref{tab:extradata}) by generating Gaussian random numbers, using the
upper and lower error bars each 50 per cent of the time.
Then we form a new sample
by culling all clusters with temperatures which fail our selection cuts
(as described in \S\ref{sec:data}).
Next we adopt a specific $M{-}T$ relation drawn
from the central value and systematic range discussed in \S\ref{sec:MT},
together with an additional scatter, different for each temperature considered,
arising from the intrinsic scatter in the $M{-}T$ relation.
For each temperature, we use the $M{-}T$ relation to calculate $n(T)$
according to the mass function in Eq.~(\ref{eqn:fnu}). 

For each cosmological model considered, we then maximize the likelihood to
find a best-fitting normalization from which to determine $\sigma_8$.
This whole process is repeated 1000 times to obtain a distribution of
$\sigma_8$ values for each cosmology.

\section{Results} \label{sec:results}

Our main result is the power spectrum normalization as a function of
cosmological model.  We quote the normalization for the variance on a
scale $R_{\rm Cl}$ that corresponds to a cluster of $6.5$ keV forming now.
This is advantageous because the value of $\si(R_{\rm Cl})$
is almost independent of the power-spectrum shape $\Gamma$
(see also Blanchard et al.~2000), while 
at fixed $\si(R_{\rm Cl})$ the value of $\si_8$ can vary by as much as 
15 per cent as $\Gamma$ spans
the 68 per cent confidence range, 0.19--0.37, of
Eisenstein \& Zaldarriaga \shortcite{EisZal}.
Fitting formulae accurate at the 1 per cent level for $R_{\rm Cl}$ and
$\si(R_{\rm Cl})$ covering the ranges
$0.2 \le \omm \le 0.8$ and $0.3 \le \oml \le 1$ are given by:
\begin{equation}
R_{\rm Cl} =p_0\ \omm^{-(p_1+p_2\,\omm+p_3\,\oml)}\,h^{-1}{\rm Mpc}\,;
\label{eqn:rn}
\end{equation}
\begin{equation}
  \sigma_{\rm Cl} \equiv
    \si(R_{\rm Cl})= p_0\ \omm^{-(p_1+p_2\,\omm+p_3\,\oml)}\,.
\label{eqn:sirn}
\end{equation}
Here $\vec p =(9.086,0.354,0.058,0.049)$
gives the coefficients in Eq.~(\ref{eqn:rn}), while
those for Eq.~(\ref{eqn:sirn}) can be found in Table~\ref{tab:cosi},
along with fits to the $\pm 1\sigma$ limits on
$\si(R_{\rm Cl})$.
The validity of Eq.~(\ref{eqn:sirn}) has been tested also in the case of 
open (i.e.~$\Omega_\Lambda=0$) and Einstein-de Sitter (i.e.~$\omm = 1$)
models, and for this wider range of parameters the
maximum deviation from the fit is still only 2 per cent.
For the concordance model ($\Omega_{\rm M}=0.3$) we find
$R_{\rm Cl}\simeq 14.8\, h^{-1}$Mpc.
A slightly less precise, but simpler fit for flat models
is given by
\begin{equation}
R_{\rm Cl}\simeq 9.0\, \Omega_{\rm M}^{-0.41}h^{-1}{\rm Mpc}.
\end{equation}
This fits to better than 1 per cent over $0.2\le\Omega_{\rm M}\le 1$.

\begin{table}
\caption{\footnotesize%
Coefficients of the fitting formula,
Eq.~(\protect\ref{eqn:sirn}), for the mean and errors of
$\si(R_{\rm Cl}) = p_0 \omm^{-(p_1+p_2\omm+p_3\oml)}$.}
\label{tab:cosi}
\begin{center}
\begin{tabular}{l|cccc}
\hline
& \multicolumn{4}{|c}{parameter} \\
& $p_0$ & $p_1$ & $p_2$ & $p_3$ \\ \hline
{\rm mean} &  $0.505$ & $0.150$  &  $-0.233$ & $0.048$ \\
$-1 \si$   &  $0.473$ & $0.124$  &  $-0.234$ & $0.053$\\
$+1 \si$   &  $0.532$ & $0.171$  &  $-0.222$ & $0.044$ \\
\hline
\end{tabular}
\end{center}
\end{table}

\begin{figure}
\begin{center}
\epsfig{file=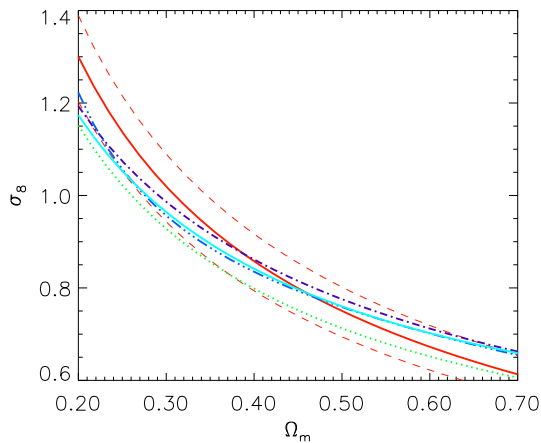 ,width=8cm}
\end{center}
\caption{The normalization $\si_8$ for flat cosmologies with shape parameter
fixed at $\Ga=0.23$
as derived here (solid,red ) with 1--$\si$ limits (dashed red lines),
compared with the fits from Viana \& Liddle \protect\shortcite{VL98}
(dot-dashed dark blue line), Wang \& Steinhardt\protect\shortcite{WanSte}
(blue 3 dots-dashed), Eke et al.~\protect\shortcite{ECF} (dotted green), and
Borgani et al.\protect\shortcite{Ste99} (cyan).}
\label{fig:s8eVL}
\end{figure}

For explicit constraints on $\sigma_8$ we verified that the following
equation is a good fit (with a maximum error of 2 per cent), in the range 
$0.19 \le \Gamma \le 0.37$ \cite{EisZal}, $0.2 \le \omm \le 0.8$ 
and $0.3 \le \oml \le 1$:
\begin{equation}
\si_8=\si_{\rm Cl}\,{q_0\, \oml^{q_1}\, \omm^{-(q_2+q_3\omm)}}\,
 \Gamma^{(q_4+q_5 \omm)}.
\label{eqn:s8}
\end{equation}
Here the vector of coefficients is given by
$\vec q =(1.049,0.029,0.546,0.350,0.278,-0.312)$.
For flat models (i.e.~$\Omega_\Lambda=1-\Omega_{\rm M}$) this corresponds
approximately to
\begin{equation}
\sigma_8\simeq (0.495^{+0.034}_{-0.037})\,\Omega_{\rm M}^{-0.60}
\end{equation}
for the value $\Gamma=0.23$ explicitly, and for general $\Gamma$:
\begin{equation}
\si_{\rm Cl}=0.462^{+0.027}_{-0.029}
\Omega_{\rm M}^{-0.194\{^{-0.179}_{-0.218}\}},
\end{equation}
where the exponents in brackets refer to the best power-laws for the
$+1\si$ and $-1\si$ limits.
It is interesting to notice that, despite the many differences in our
approach, this result does not differ dramatically from other studies,
such as those reported in Fig.~\ref{fig:s8eVL}.  This is true even
for those studies with apparently quite
discrepant temperature functions, such as Henry~\shortcite{Henry}.  As
we discuss below this is partly due to the chance cancelation of several small
changes affecting the derived $\sigma_8$.

We performed several checks on our results.  First we made sure that
analysing the sample using a single (median) redshift did not introduce a
significant bias.  We split the sample into low-$z$ and high-$z$ halves and
analysed them independently.  The variation in $\si_{\rm Cl}$ between the
low-$z$, high-$z$ and combined samples was only 2 per cent for a flat model
with $\omm=0.3$.  We also gradually eliminated clusters with redshift below a
threshold $z_*$, and then
above that threshold. The results change smoothly with $z_*$, implying that
the final result is not dominated by any particular cluster.
Finally we eliminated all clusters below $6\,$keV (which approximately halves
the sample in each Monte-Carlo realization).  While the errors increase
in this process, the central value only increases by 3--5 per cent
(depending on $\omm$).

In earlier cluster samples the shape of the temperature function was not
obviously  well fit by the theoretical predictions.
Much of this discrepancy has now disappeared
thanks to better data.  The question remains however as to how much the shape
of the temperature function affects the fit, rather than for example the
overall amplitude at some intermediate temperature.  To address this we have
calculated $\si_{\rm Cl}$ for flat cosmologies by matching 
the observed $n({>}\,kT_{\rm X})$ for $T_{\rm X}=6\,$keV.
We found values 2.5--4.5 per cent lower than fitting to the whole
temperature function for $T_{\rm X}>3.5\,$keV. 
Thus the shape of the temperature function is being used in the fit, and
shifts the best fitting $\si_{\rm Cl}$ by a small, but non-negligible amount.

In Fig.~\ref{fig:s8eVL} we compare our results with those of
Viana \& Liddle~\shortcite{VL98} and others for flat cosmologies as a function
of $\omm$.\footnote{Many authors have found similar results which
slightly differ from Viana \& Liddle.  We choose to focus on their study for
comparison, since they were very explicit about the details of their procedure.
See Wang \& Steinhardt \shortcite{WanSte} for a detailed discussion of
differences between some other studies.}
While the Viana \& Liddle \shortcite{VL98}
best fit is within our error bars, the shape 
of the two curves is quite different.
This discrepancy can be traced to a number of factors:
a newer data compilation; our likelihood fit to the entire cluster data;
our use of the Sheth \& Tormen \shortcite{SheTor} universal mass function
rather than the PS theory; not integrating over formation redshifts;
different $M{-}T$ normalization and scatter;
and different sources of errors (in particular Viana \& Liddle~1999
included errors in $\Gamma$).

\begin{figure}
\begin{center}
\epsfig{file=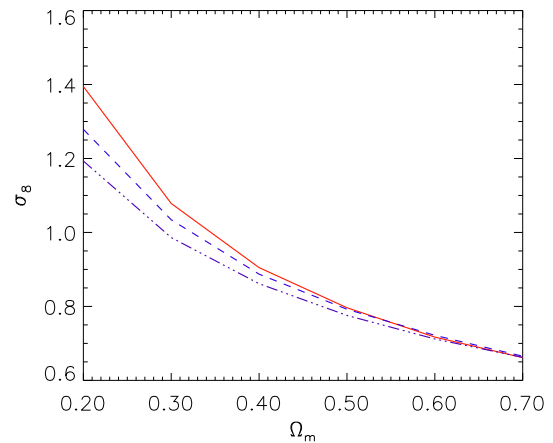 ,width=8cm}
\end{center}
\caption{The normalization $\si_8$ for flat cosmologies with $\Ga=0.23$
as derived here (solid red line) but with $\beta = 1.17$,
compared with the Viana \& Liddle \protect\shortcite{VL98} results 
(dark blue dot-dashed line) and also compared with their approach but without
the integration over formation redshift (blue dashed line).}
\label{fig:s8_fl_cVLnoz}
\end{figure}

We found a relatively symmetric error bar on our final results
for $\sigma_{\rm Cl}$ and $\sigma_8$, while Viana \& Liddle \shortcite{VL98}
had a very asymmetric error.  It appears that this is mainly due to their
assumed skew distribution for $\Gamma$.  Another important difference is that
Viana \& Liddle \shortcite{VL98}
adopted a lower value of $\beta$ with respect to ours.
If we chose $\beta = 1.17$, we would find the same normalization 
at high $\omm$ (see Fig.~\ref{fig:s8_fl_cVLnoz}). 
Approximately half of the discrepancy at lower $\omm$  comes from the
fact that we have not integrated over `formation' redshift, while
Viana \& Liddle~\shortcite{VL98} included such an integration
 (see Fig.~\ref{fig:s8_fl_cVLnoz}).
We argued in \S\ref{sec:theory} that including such an integration in addition
to the scatter in the $M{-}T$ relation was effectively double counting
some of the scatter.

Use of the Sheth \& Tormen \shortcite{SheTor} or
Jenkins et al.~\shortcite{JFWCCEY} mass function,
rather than PS, {\it lowers\/} our
results by 4--8  per cent (depending on $\omm$). Note that the overall
effect of the different mass function may depend on redshift, so that if
formation redshift is integrated over then the discrepancy may even 
be in the other direction.

\begin{figure}
\begin{center}
\epsfig{file=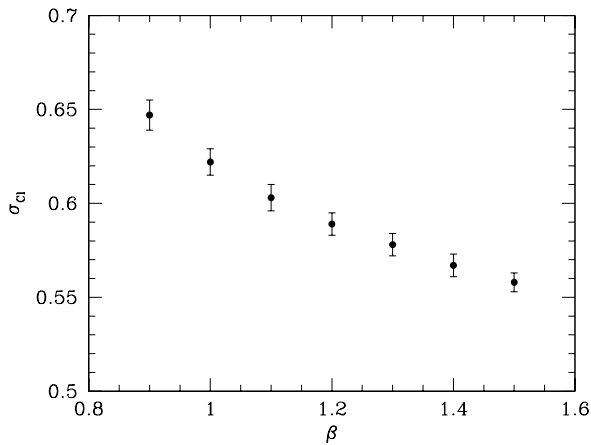 ,width=8cm}
\end{center}
\caption{The largely $\Gamma$-independent normalization $\si(R_{\rm Cl})$
for different values of $\beta$ in the $M{-}T$ relation.  The error bars
indicate the standard deviation of $\sigma(R_{\rm Cl})$ arising from all
uncertainties except for the systematic error in $\beta$, which is set to
zero here.  Our adopted central value of $\beta$ is 1.3.}
\label{fig:siga}
\end{figure}

A major source of theoretical uncertainty remains the value and distribution
of $\beta$ in the $M{-}T$ relation of Eq.~(\ref{eqn:m-t}).
We show in Fig.~\ref{fig:siga} how $\si_{\rm Cl}$ changes as $\beta$ is
scanned {}from 0.9 to 1.5 (with no systematic error) in our reference model
with $\omm=0.3$ and $\oml=0.7$. 
It is clear that a better estimate of the $M\,{-}\,T$ relation would 
greatly reduce the errors on $\si_{\rm Cl}$.  It is also clear that to obtain
an unbiased estimate of $\si_{\rm Cl}$ it is essential to use an appropriate
central value and error on $\beta$, which is why we have included both a
statistical and `systematic' error in our fits (see also Table~6).

An estimate of the uncertainty introduced by various factors is shown in
Table~\ref{tab:unc}.
To obtain the estimates of error budget in this table we ran our code
with the sources of errors described in the first column.
The first case for example is the complete calculation, with all
sources of error included.
The second case has only the statistical uncertainty in the $M{-}T$ relation,
while the uncertainty in the value of the cluster temperature and the
systematic scatter in the $M{-}T$ relation were not used.  The other cases
show what happens with one or other source of uncertainty excluded.
The mean quoted is the mean of the distribution obtained in each case,
while the the errors are obtained by integrating the normalized distribution
until 34 per cent of the area is reached on each side of the mean.

We find that our uncertainty in $\sigma_8$ is dominated by systematics in
the $M{-}T$ relation (see also Voit~2000).
The statistical scatter in the $M{-}T$ relation has an almost negligible
effect on the central value and the error.
The temperature errors themselves give a skewed distribution to $\sigma_8$,
but this effect is sub-dominant.

\section{Conclusions} \label{sec:conclusions}

\begin{table}
\caption{\footnotesize%
For a flat cosmology with $\omm = 0.3$ we estimate how each source 
of uncertainty affects the mean and the error of our $\sigma_{\rm Cl}$
estimates.  $M{-}T|_{\rm sys}$ and
$M{-}T|_{\rm stat}$ represent the systematic and statistical
errors in the $M{-}T$ relation, while $T$ represents the error in the 
temperature of the clusters.  The $+1\si$ and $-1\si$ limits are found
from integrating the distribution of 1000 Monte Carlos in each case.} 
\label{tab:unc}
\begin{center}
\begin{tabular}{l|ccc}
\hline
Source of error           & mean $\si(R_{\rm Cl})$  & $+1\si$ & $-1\si$ \\
\hline
all & 0.581 & 0.049 & 0.050 \\
$M{-}T|_{\rm sys}$                    & 0.575 & 0.047 & 0.047 \\
$M{-}T|_{\rm stat}$                   & 0.570 & 0.002 & 0.002 \\
$T$                                  & 0.586 & 0.005 & 0.018 \\
\hline
\end{tabular}
\end{center}
\end{table}

The local abundance of rich clusters of galaxies currently provides one of
the strongest constraints on the normalization of the present day dark matter
power spectrum on a scale of ${\sim}\,10\,$Mpc.
While there have been numerous detailed
studies of this constraint in the past, recent developments in both theory
and observation have made it worthwhile to revisit this quantity.

We have calculated the constraint on $\sigma_8$ arising from a new local sample
of X-ray clusters with \ASCA\ temperatures.  We have incorporated the universal
mass function determined from recent large N-body simulations, which is
sufficiently different from the PS theory that the cosmological constraints
inferred from cluster abundances change.
We tried to carefully define the relation between mass and X-ray temperature
for galaxy clusters, based on the results of many different hydrodynamical
simulations.  Another major difference with previous work was that we
performed all the comparisons at the `observed epoch' rather than carrying
out an integration over `formation times'.  We also considered general
combinations of $\Omega_{\rm M}$ and $\Omega_{\rm \Lambda}$, including closed
models.  Our results are best presented in terms of a normalization at the
characteristic scale for our sample, $R_{\rm Cl}$, where the normalization is
largely independent of the power spectrum shape.

In the near future we anticipate a dramatic increase in our knowledge of the
cosmological parameters from CMB anisotropy missions.  In particular it has
been forecast \cite{EHT} that \MAP\ will be
able to determine $\sigma_8$ to 14 per cent, at the same time as constraining a
suite of other parameters.  The cluster abundance constraint is currently
uncertain only at the ${\sim}\,10$ per cent level, and so certainly is a
pivotal limit on models -- it is also an entirely independent constraint
(at low $z$, in the mildly non-linear regime) compared with the
CMB determinations (derived from the purely linear regime at $z\simeq1000$).
The uncertainty in the derived $\sigma_8$ could be dramatically
reduced with improvement in the $M{-}T$ relation, as well as through new
X-ray data coming from \Chandra\ and \XMM.

Future X-ray surveys, which go much fainter will lead to an increase in the
size and quality of the X-ray data.  We showed in Fig.~\ref{fig:vT}
that there is lots of volume that was not probed at, say, $5\,$keV.
For example the \ROSAT\ Bright Survey \cite{RBS} contains an approximately
complete flux-limited sample of 302 clusters at $|b|\,{>}\,30^\circ$, most
of them lying at low redshift, which would be an excellent data-base if they
all had good temperature estimates.
More important than just increasing the precision of the X-ray temperature
measurements is an increase in the quality of data, for the purposes of more
fully understanding the comparison with models.
Better spatial and spectral information from X-ray clusters should allow
more reliable estimates for the value of $T_{\rm X}$ which is most
appropriate for comparing with the simulations.  In a similar vein,
improved optical studies of lensing and velocity dispersions could improve
mass determinations for individual clusters.

The major improvement will come through a more precise mass-temperature
relation for X-ray clusters.
This will require both the improved data which we can foresee from
\Chandra\ and \XMM, {\it and\/} improvements in modelling the gas physics
relevant for understanding the X-ray properties of clusters.
Eventually we imagine that the data will be fit more directly to much more
ambitious simulations -- this will be extremely difficult in practice, since
a wide range of scales needs to be modelled simultaneously.
In the meantime the piece-meal approach we have taken here will continue to be
useful.  With each of the ingredients, particularly the
mass-temperature relation, improving with time, the cluster abundance
will continue to be a strong constraint on the normalization of dark matter
fluctuations on the ${\sim}\,10\,$Mpc scale.

\subsection*{ACKNOWLEDGMENTS}
EP and DS were supported by the Canadian Natural Sciences and Engineering
Research Council, and MW by the US National Science Foundation and a Sloan
Fellowship.
EP is a National Fellow of the Canadian Institute for Theoretical
Astrophysics.
We would like to thank the Institute for Theoretical Physics, Santa
Barbara for their hospitality while some of this work was carried out.
We are indebted to
Alain Blanchard, Stefano Borgani, Patrick Henry, Andrew Liddle,
Maxim Markevitch and David Weinberg for providing useful comments on an
early version of the manuscript.
This research has made use of the NASA/IPAC Extragalactic Database (NED)
which is operated by the Jet Propulsion Laboratory,
California Institute of Technology, under contract with the
National Aeronautics and Space Administration.

\end{document}